\documentclass[letterpaper, 10 pt, conference]{ieeeconf}  

\IEEEoverridecommandlockouts                              

\usepackage{graphicx} 
\usepackage{amsmath} 
\usepackage{amssymb}  
\usepackage{hyperref}
\usepackage{soul}
\usepackage{float}
\usepackage{bm}
\usepackage{cite}

  \usepackage{amsthm}
\theoremstyle{definition}

\theoremstyle{definition}

\theoremstyle{plain}
\newtheorem{theorem}{Theorem}
\theoremstyle{plain}

\theoremstyle{definition}
\theoremstyle{definition}

\theoremstyle{definition}

\newtheorem{remark}{Remark}

\usepackage[table,xcdraw]{xcolor}
\usepackage{easyReview}

\title{\LARGE \bf
Zero Dynamics Stability of a Multirotor UAV with a Heavy Payload
}

\author{Sander Doodeman$^{1}$, Paula Chanfreut$^{1}$, Elena Torta$^{1}$, Duarte Antunes$^{1}$\\
\thanks{$^{1}$ Department of Mechanical Engineering, TU/e, Eindhoven University of Technology, Eindhoven, The Netherlands}%
\thanks{Email: {\tt s.doodeman@tue.nl}}%
\thanks{This research is supported by the Renewable Energy Transition Topsector Energy subsidies (HER+22-02-03430912) from the Dutch Ministry of Economic Affairs.}%
}

\begin{document}

\maketitle
\thispagestyle{empty}
\pagestyle{empty}

\begin{abstract}
This paper studies the zero dynamics stability of a multirotor Unmanned Aerial Vehicle (UAV) carrying a rigidly attached heavy payload.
In particular, we focus on controlling the position of a virtual task-relevant point of interest (\textit{POI}) on the UAV and investigate how its stability depends on payload placement.
Taking the \textit{POI} position and the yaw angle as output, we prove that the zero dynamics are Lyapunov stable only if this \textit{POI} is above the system's total center of gravity, assuming that the \textit{POI} and the payload are both on the UAV's body-fixed $z$-axis.
Remarkably, this indicates that when the payload coincides with the \textit{POI}, zero dynamics stability is only guaranteed if the payload is above the UAV rather than below it.
For the more general case where the \textit{POI} and payload do not necessarily lie on the same $z$-axis, we demonstrate stability for a simplified planar model.
The analysis also yields an explicit bound on the horizontal payload offset given the rotor thrust limits.
Finally, numerical MATLAB simulations provide insight into how results for the planar model are also observed for the general 3D system.
\end{abstract}

\section{INTRODUCTION}
Advances in control techniques have enabled multirotor UAVs to autonomously perform increasingly complex tasks that would otherwise be time-consuming or expensive.
This includes, for example, camera- and X-ray-based inspections \cite{meere_x-ray_2025, he_uav_2025} and window cleaning tasks \cite{hulaibi_skybot_2023}, all requiring heavy payloads on a UAV.
With increasing duration of these tasks, larger and heavier batteries are required, adding additional weight.
When designing task-specific UAVs, using off-the-shelf UAV components is often preferred.
However, the added equipment can significantly affect the UAV's total center of gravity (\textit{CoG}) and closed-loop behavior.

Considering the position of a rigidly attached payload (\textit{PL}), it is beneficial to horizontally balance the payload near the center of the UAV with respect to the rotors.
Otherwise, a steady-state torque needs to be applied, leading to reduced efficiency.
However, the effect of the vertical placement (above or below the UAV) of the payload is more subtle.
For this, the task at hand should have priority.
In fact, rather than controlling the center point of the UAV, the position of the payload or any other virtual point of interest (\textit{POI}) on the UAV should be accurately controlled.
This is especially relevant when a specific sensor on a UAV has to be stabilized, e.g., for camera-based inspections.
However, most works in literature consider the control of the \textit{CoG}, rather than the control of the task-defined \textit{POI}.
This applies even to works that consider the control of the tandem between the multirotor UAV and a heavy payload \cite{molina_dynamic_2017, raina_automated_2022}.

One frameworks for addressing the effect of the payload position (or another \textit{POI}) on the stability and performance of the system is a zero dynamics analysis \cite{nguyen_mechanics_2015}.
Zero dynamics describe the internal behavior of a system when the output (in this case, the \textit{POI}) is constrained to remain identically zero. Their stability reveals whether the system is minimum phase and provides a fundamental measure of its inherent internal stability \cite{isidori_zero_2013}.

\par Nguyen et al. have analyzed zero dynamics when the controlled outputs are the $x$-, $y$-, $z$-positions and the yaw of the \textit{POI} \cite{nguyen_mechanics_2015}.
When these variables are regulated to zero, the remaining zero dynamics for the roll, pitch, yaw, roll rate, and pitch rate of the UAV are to be analyzed.
An important assumption in \cite{nguyen_mechanics_2015} is that the payload mass is negligible, and the analysis is limited to the linearized zero dynamics around the hovering equilibrium.
It is proven that the linearized zero dynamics are unstable when the \textit{POI} is placed below the \textit{CoG}, which implies instability of the zero dynamics.
The intuition behind this is that when the \textit{POI} is below the \textit{CoG} and the UAV is slightly tilted, the thrust vector causes the UAV to tilt even more.
However, when the \textit{POI} is placed above the \textit{CoG}, \cite{nguyen_mechanics_2015} shows that the eigenvalues of the linearized dynamics are on the imaginary axis.
Thus, no conclusion about (marginal) stability for this case can be drawn \cite{khalil_nonlinear_2002}.
On top of this, previous work on the stability of a multirotor UAV with a payload does not consider individual motor thrust limits \cite{nguyen_mechanics_2015}.

\par This paper provides new insight into the stability of the zero dynamics of a multirotor UAV with a heavy payload by tackling the problem left open in \cite{nguyen_mechanics_2015}.
Namely, we prove Lyapunov stability for the nonlinear zero dynamics of the full multirotor UAV model.
For this, we assume that the heavy payload and \textit{POI} to be controlled are on the body-fixed $z$-axis of the UAV.
Our proof relies on providing an explicit Lyapunov function.
For the more general case where the \textit{PL} and \textit{POI} are not necessarily on the $z$-axis of the UAV, we consider a planar model of the UAV and prove Lyapunov stability.
Moreover, we consider the effect of thrust limits on the stability of the planar system and how this is affected by the horizontal position of the \textit{PL}.
We provide a numerical analysis for the 3D system with a horizontally off-centered \textit{PL} and \textit{POI} position.
The drawn conclusions can be used as a design guide for placing heavy payloads and task-critical sensors on multirotor UAVs.

\par The remainder of the paper is organized as follows.
Section \ref{sec:preliminaries} provides some preliminaries and establishes notation.
Section \ref{sec:3D_proof} briefly discusses the model of the multirotor UAV with heavy payload, derives the zero dynamics, and provides our main results on the stability of the zero dynamics.
Section \ref{sec:planar_proof} discusses the planar case.
Simulation results are provided in Section \ref{sec:simulations} and concluding remarks in Section \ref{sec:conclusion}.



\section{PRELIMINARIES} \label{sec:preliminaries}
\begin{figure}
    \centering
    \includegraphics[width=0.65\linewidth]{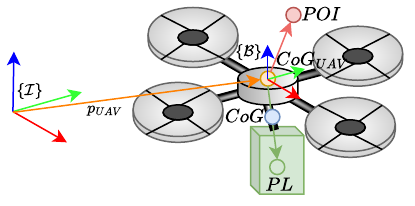}
    \vspace{-5pt}
    \caption{Relevant points and coordinate frames of the multirotor UAV}
    \label{fig:3D_UAV_points}
    \vspace{-10pt}
\end{figure}

Let us define $g=9.81\,[m/s^2]$ as the gravity acceleration, $I_n$ as the $n\times n$ identity matrix, $\bm{e}_3 = \begin{bmatrix}
    0 & 0 & 1
\end{bmatrix}^\top$ as the third unit vector, and $S(.)$ as the skew-symmetric matrix operator such that $S(a)b = a \times b$.
Let $m_{UAV}$ and $m_{PL}$ denote the mass of the UAV and rigidly attached payload defined in the body-fixed frame, respectively, and let $m_{\text{tot}}$ be the total mass of the system (UAV and payload).
Also, let $H_{UAV}$ and $H_{PL}$ be the inertia of the UAV and payload expressed in the body-fixed frame, respectively, and let $H_{\text{tot}} = H_{UAV} + H_{PL}$.
Furthermore, let $\bm{\eta}$ be the $ZYX$ Euler angle representation of the rotation of the UAV with respect to the inertial frame, composed as $\bm{\eta} = \begin{bmatrix}
    \varphi & \theta & \psi
\end{bmatrix}^\top$.
From $\bm{\eta}$, a rotation matrix $R$ can be obtained, which maps vectors in the body-fixed frame of the UAV to the inertial frame.
$\bm{\omega}$ is the angular velocity of the UAV with respect to the body-fixed frame.
The relation between $\bm{\omega}$ and $\dot{\bm{\eta}}$ is given by
\begin{equation*}
    \bm{\omega} = \Gamma \dot{\bm{\eta}}, \quad \Gamma=\begin{bmatrix}
        1 & 0 & -\sin(\theta)\\
        0 & \cos(\varphi) & \sin(\varphi)\cos(\theta)\\
        0 & -\sin(\varphi) & \cos(\varphi)\cos(\theta)
    \end{bmatrix}.
\end{equation*}
Let $\bm{p}_{UAV}$ be the position of the center of gravity of the UAV without the payload (\textit{CoG}$_{UAV}$) expressed in the inertial frame, as shown in Fig. \ref{fig:3D_UAV_points}.
The positions of the \textit{POI} and the \textit{PL} in the body-fixed frame are denoted by
\begin{equation*}
    \bm{r}_{POI} = \begin{bmatrix}
        x_{POI} \\ y_{POI} \\ z_{POI}
    \end{bmatrix}, \quad \bm{r}_{PL} = \begin{bmatrix}
        x_{PL} \\ y_{PL} \\ z_{PL}
    \end{bmatrix}.
\end{equation*}
In the inertial frame, the positions of the \textit{POI} ($\bm{p}_{POI}$) and the payload ($\bm{p}_{PL}$) are related by $\bm{p}_{POI} = \bm{p}_{UAV} + R\bm{r}_{POI}$, $\bm{p}_{PL} = \bm{p}_{UAV} + R\bm{r}_{PL}$.
Define $\hat{\bm{r}}$ as $\hat{\bm{r}} = \bm{r}_{POI} - \frac{m_{PL}}{m_{\text{tot}}}\bm{r}_{PL}$,
representing the position of the \textit{POI} relative to the total \textit{CoG} of the system, such that $\bm{p}_{CoG} = \bm{p}_{POI} - R\hat{\bm{r}}$.
Finally, the control input $\bm{F}_u$ is defined as $\bm{F}_u = \begin{bmatrix}
    T & \bm{\tau}^\top
\end{bmatrix}^\top$, where $T$ and $\bm{\tau}$ are the total thrust and the torque of the UAV, respectively.


\section{LYAPUNOV STABILITY FOR 3D MULTIROTOR UAV WITH HEAVY PAYLOAD} \label{sec:3D_proof}
In this section, we analyze the zero dynamics of the multirotor UAV system with a heavy payload when the \textit{POI} and \textit{PL} are on the body-fixed $z$-axis of the UAV.
For this, given the state $\bm{x} = \begin{bmatrix}
    \dot{\bm{p}}_{POI}^\top & \bm{\omega}^\top & \bm{p}_{POI}^\top & \bm{\eta}^\top
\end{bmatrix}^\top$, we analyze the dynamic equations of the system.
For this, we define the kinetic energy $\mathcal{K}$ and potential energy $\mathcal{U}$ as 
\begin{gather}
\begin{aligned}
    \mathcal{K} &= \frac{1}{2} m_{\text{tot}} \dot{\bm{p}}_{CoG}^\top\dot{\bm{p}}_{CoG} + \frac{1}{2} \bm{\omega}^\top H_{CoG}(\bm{r}_{PL}) \bm{\omega}, \\
    \mathcal{U} &= m_{\text{tot}} g \bm{e}_3^\top \bm{p}_{CoG},
\end{aligned}
\end{gather}
where $\bm{g} = g\bm{e}_3$ and $H_{CoG}$ is given by 
\begin{equation}
    H_{CoG}(\bm{r}_{PL}) = \frac{m_{UAV} m_{PL}}{m_\text{tot}} S(\bm{r}_{PL})^\top S(\bm{r}_{PL}) + H_\text{tot}.
\end{equation}
Then, using a Lagrangian approach \cite{kraker_mechanical_2001, lippiello_cartesian_2012}, the dynamics of the system can be written as
\begin{gather}
\label{eq:dyn_eq_3D}
\begin{aligned}
    &\dot{\bm{x}} = f(\bm{x}, \bm{F}_{u})= \\
    & \begin{bmatrix}
        B_{y}(\bm{\eta})^{-1} (M_F \bm{F}_u - C_{y}(\dot{\bm{p}}_{POI}, \bm{\omega}, \bm{\eta}) \begin{bmatrix}
            \dot{\bm{p}}_{POI} \\
            \bm{\omega}
        \end{bmatrix} - G_{y}(\bm{\eta})) \\
        \begin{bmatrix}
            I_3 & 0_{3\times 3} \\ 0_{3\times 3} & \Gamma^{-1}
        \end{bmatrix} \begin{bmatrix}
            \dot{\bm{p}}_{POI} \\
            \bm{\omega}
        \end{bmatrix}
    \end{bmatrix},
\end{aligned}
\end{gather}
in which
\begin{gather*}
    B_{y}(\bm{\eta}) = \begin{bmatrix}
        m_{\text{tot}}I_3 & m_{\text{tot}} R S(\hat{\bm{r}}) \\
        m_{\text{tot}} S(\hat{\bm{r}})^\top R^\top & M_1
    \end{bmatrix}, \\
    C_{y}(\dot{\bm{p}}_{POI}, \bm{\omega}, \bm{\eta}) = \begin{bmatrix}
        0_{3\times 3} & m_{\text{tot}} R S(\bm{\omega}) S(\hat{\bm{r}}) \\
        m_{tot}M_2 R^\top & S(\bm{\omega}) M_1 - M_3
    \end{bmatrix}, \\
\begin{aligned}
    G_{y}(\bm{\eta}) =& \begin{bmatrix}
        m_{\text{tot}} g \bm{e}_3 \\
        m_{\text{tot}} g S(\hat{\bm{r}}) \Gamma e_3
    \end{bmatrix}, \quad
    M_F = \begin{bmatrix}
        R \bm{e}_3 & 0_{3\times 3} \\ \bm{0} & I_3
    \end{bmatrix},\\
    M_1 =& H_{CoG}(\bm{r}_{PL}) + m_\text{tot}S(\hat{\bm{r}})^\top S(\hat{\bm{r}}), \\
    M_2 =& S(\hat{\bm{r}}) S(\bm{\omega}) - S(\bm{\omega})S(\hat{\bm{r}}), \\
    M_3 =& m_{\text{tot}} ({\Gamma^{-1}}^\top \otimes \dot{\bm{p}}_{POI}^\top) R_{\bm{\eta}} S(\hat{\bm{r}}), \\
    R_{\bm{\eta}} =& \begin{bmatrix}
        \frac{\partial R^\top}{\partial \varphi} & \frac{\partial R^\top}{\partial \theta} &  \frac{\partial R^\top}{\partial \psi}
    \end{bmatrix}^\top.
\end{aligned}
\end{gather*}
\subsection{Zero Dynamics}
Consider the output of the system to be $\bm{y} = \begin{bmatrix}
    x_{POI} &
    y_{POI} &
    z_{POI} &
    \psi
\end{bmatrix}^\top$.
Assume that both \textit{POI} and \textit{PL} lie on the body-fixed $z$-axis of the UAV (such that $x_{POI} = y_{POI} = x_{PL} = y_{PL} = 0$).
The body-fixed frame is assumed to coincide with the principal axes of inertia, resulting in a diagonal total inertia matrix $H_{\text{tot}} = \texttt{diag}(h_{11}, h_{22}, h_{33})$.
This is justified by the approximately symmetric mass distribution of the UAV without payload.
Performing an input-output linearization to derive the zero dynamics yields the control law under which the resulting behavior corresponds to the zero dynamics
\begin{equation}
\label{eq:zd_control_law}
    \bm{F}_{u,\text{zd}} = \begin{bmatrix}
        m_{\text{tot}} \left(\alpha (\dot{\varphi}^2 + \cos(\varphi)^2\dot{\theta}^2) + g\cos(\varphi)\cos(\theta)\right) \\
        -\frac{1}{2}\sigma_1\sin(2\varphi)\dot{\theta}^2 - \frac{g}{\alpha}\sin(\varphi)\cos(\theta)\left(h_{11} + \sigma_3\right) \\
        \sigma_2\sin(\varphi)\dot{\varphi}\dot{\theta} - \frac{g}{\alpha}\sin(\theta)\left(h_{22} + \sigma_3\right) \\
        \sigma_2 \cos(\varphi) \dot{\varphi} \dot{\theta} - \frac{2 h_{33}}{\cos{\varphi}}\left( \dot{\varphi} \dot{\theta} - \frac{g}{\alpha} \sin(\varphi) \sin(\theta) \right)
    \end{bmatrix},
\end{equation}
where 
\begin{gather*}
    \alpha = z_{POI} - \frac{m_{PL}}{m_{\text{tot}}} z_{PL} \neq 0, \quad \sigma_1 = h_{11} - h_{22} + h_{33}, \\
    \sigma_2 = -h_{11} + h_{22} + h_{33}, \quad
    \sigma_3 = \frac{m_{PL}m_{UAV}}{m_{\text{tot}}}z_{PL}^2.
\end{gather*}
Here, $\alpha$ represents the vertical distance between the \textit{POI} and the \textit{CoG}.
\begin{remark}
    It is well-known that when $\alpha=0$, i.e., when the \textit{CoG} coincides with the \textit{POI}, the system is differentially flat and thus the zero dynamics are trivial \cite{mellinger_minimum_2011}.
    Therefore, we are only interested in the cases $\alpha>0$ and $\alpha<0$.
\end{remark}
After replacing \eqref{eq:zd_control_law} in \eqref{eq:dyn_eq_3D}, and after extensive derivations, this results in the following relations for the zero dynamics for state $\hat{\bm{x}} = \begin{bmatrix}
    \dot{\varphi} &
    \dot{\theta} &
    \varphi &
    \theta
\end{bmatrix}^\top$, given that $\alpha \neq 0$:
\begin{gather}
\label{eq:zerodyn}
    \dot{\hat{\bm{x}}} =
    \begin{bmatrix}
        \ddot{\varphi} \\
        \ddot{\theta} \\
        \dot{\varphi} \\
        \dot{\theta}
    \end{bmatrix} = \begin{bmatrix}
        -\frac{\sin(\varphi) (\alpha \cos(\varphi) \dot{\theta}^2 + g \cos(\theta))}{\alpha} \\
        -\frac{g \sin(\theta) - 2 \dot{\varphi} \dot{\theta} \alpha \sin(\varphi)}{\alpha \cos(\varphi)} \\
        \dot{\varphi} \\
        \dot{\theta}
    \end{bmatrix}.
\end{gather}
\par The next result is the main result of the paper, establishing that the zero dynamics \eqref{eq:zerodyn} are Lyapunov stable when $\alpha > 0$.
For this, we present an invariant Lyapunov function to show that solutions to the zero dynamics in \eqref{eq:zerodyn} are bounded.
Therefore, let us define the region $\mathcal{D} = \left\{\hat{\bm{x}} \in \mathbb{R}^4 : |\varphi| < \frac{\pi}{2}, |\theta| < \frac{\pi}{2}\right\}$.
Then, considering the Lyapunov function candidate
\begin{equation} \label{eq:lyapfcn}
    V(\hat{\bm{x}}) = \frac{\alpha}{2g}\left( \dot{\varphi}^2 + \cos(\varphi)^2\dot{\theta}^2 \right) + \left( 1 - \cos(\varphi)\cos(\theta) \right),
\end{equation}
let us define the set $\Omega_c = \{ \hat{\bm{x}} : V(\hat{\bm{x}}) \leq c \}$ for any $c$ such that $0 \leq c < 1$.
Finally, we define $\bar{\Omega}_c$ as the intersection of these sets: $\bar{\Omega}_c = \mathcal{D} \cap \Omega_c$.
\begin{theorem}
     Let $c \in \mathbb{R}$, $0\leq c<1$. If $\alpha > 0$, then 
  \begin{itemize}
      \item[(i)] $\hat{\bm{x}}(t) \in \bar{\Omega}_c$ for all $t \geq 0$ when $\hat{\bm{x}}_0 \in \bar{\Omega}_c$;
      \item[(ii)] the zero dynamics given in \eqref{eq:zerodyn} are Lyapunov stable around the origin.
  \end{itemize}  
  Moreover, if $\alpha < 0$, the zero dynamics in \eqref{eq:zerodyn} are unstable around the origin.
\end{theorem}
\begin{proof}
    Let us consider the Lyapunov function candidate in \eqref{eq:lyapfcn}, which is continuously differentiable on $\mathcal{D}$, a set containing the origin.
    Furthermore, $V(\hat{\bm{x}}) \geq 0$ for $\alpha > 0$, and $V(\hat{\bm{x}})=0$ only for $\hat{\bm{x}}=0$, indicating positive definiteness of $V(\hat{\bm{x}})$.
    Substituting the dynamics in \eqref{eq:zerodyn}, the time derivative of this Lyapunov function candidate is given by
    \begin{equation*}
    \begin{aligned}
        \dot{V}(\hat{\bm{x}}) =& \frac{\alpha}{g}\left( \dot{\varphi}\ddot{\varphi} + \sin(\varphi)\cos(\varphi)\dot{\varphi}\dot{\theta}^2 + \cos(\varphi)^2\dot{\theta}\ddot{\theta} \right) \\
        &+ \sin(\varphi)\cos(\theta)\dot{\varphi} + \cos(\varphi)\sin(\theta)\dot{\theta} =0
    \end{aligned}
    \end{equation*}
    Suppose $\hat{\bm{x}}(0) \in \Omega_c$. Then $V(\hat{\bm{x}}(0))
    \leq c$ and since the derivative of $V$ along systems trajectories is zero, we have $V(\hat{\bm{x}}(t))=V(\hat{\bm{x}}(0))
    \leq c$ and thus $\hat{\bm{x}}(t)\in \Omega_c$ for all $t \geq 0$.  
    Even more, any initial state in $\bar{\Omega}_c$ cannot leave $\bar{\Omega}_c$, which we prove by contradiction.
    Suppose a solution crosses the boundary of $\mathcal{D}$.
    On this set boundary, according to the set definition, $\cos(\varphi)=0$ or $\cos(\theta)=0$.
    Then, according to \eqref{eq:lyapfcn}, $V(\hat{\bm{x}}) = \frac{\alpha}{2g}\left( \dot{\varphi}^2 + \cos(\varphi)^2\dot{\theta}^2 \right) + 1 \geq 1 > c$ for $\alpha > 0$.
    Since $\hat{\bm{x}}(t) \in \Omega_c$ for all $t \geq 0$ this leads to a contradiction.
    Hence, if $\hat{\bm{x}}(0) \in \bar{\Omega}_c$, $\hat{\bm{x}}(t) \in \bar{\Omega}_c$ for all $t \geq 0$ if $\alpha > 0$. This proves (i).
    To prove (ii), we use Lyapunov's stability \mbox{theorem~\cite[Th. 4.1]{khalil_nonlinear_2002}}.
    $\dot{V}(\hat{\bm{x}}) = 0$ indicates that the Lyapunov function $V(\hat{\bm{x}})$ is an invariant function, which is constant along the system trajectories, bounding the trajectories to its level sets and proving Lyapunov stability \cite{khalil_nonlinear_2002}.
    
    Finally, whenever \hbox{$\alpha<0$}, linearizing the system around the hovering equilibrium point results in
    \begin{equation*}
        \dot{\hat{\bm{x}}} = A\hat{\bm{x}}, \quad A = \begin{bmatrix}
            0_{2\times2} & -\frac{g}{\alpha} I_2 \\
            I_2 & 0_{2\times2}
        \end{bmatrix}.
    \end{equation*}
    Since the eigenvalues of $A$ are given by $0$ and $\pm\sqrt{-\frac{g}{\alpha}}$, there are two positive eigenvalues of $A$ for \hbox{$\alpha<0$}, indicating that the system is unstable according to Theorem 4.7 in \cite{khalil_nonlinear_2002}.
\end{proof}

Proving that the origin is stable when the \textit{POI} is above the total \textit{CoG} of the UAV and does not necessarily lie on the body-fixed $z$-axis of the UAV is intractable and out of the scope of this paper.
However, as we show in the next section, we can pursue such an analysis for a simplified planar model.

\begin{remark}
    In this zero dynamics analysis, no thrust limits of the UAV rotors have been taken into account to generalize the result for different multirotor UAVs.
    Most UAVs do not allow negative motor thrusts; hence, it is important to check the impact of thrust saturation when assessing stability of the zero dynamics.
    From the control law \eqref{eq:zd_control_law}, we can derive that when the \textit{POI} is above the \textit{CoG} ($\alpha > 0$), the total thrust value $T$ cannot be negative as long as $|\varphi|,|\theta| < \frac{\pi}{2}$.
    This is not true for $\alpha < 0$.
    More insight into the effect of thrust limits will be given for the planar model in the next section.
\end{remark}

\section{LYAPUNOV STABILITY FOR PLANAR MULTIROTOR UAV WITH PAYLOAD} \label{sec:planar_proof}
In this section, we investigate the role of the horizontal placement of the payload on the stability of the zero dynamics.
We do this for a planar representation of the system in the $(y,z)$-plane to enable a more tractable analysis.
In Section \ref{sec:4_modeling}, we derive the dynamic model for this planar system representation.
A new feature arises when we consider non-zero $x_{PL}$ and $y_{PL}$, namely, we need to derive the maximum distance of the payload position with respect to the body-fixed $z$-axis of the UAV, which is limited when thrust limits are applied.
We carry out this analysis in Section \ref{sec:4_thrustlimits}.
In Section \ref{sec:4_stability}, we prove that the zero dynamics of the planar system are stable even when the \textit{POI} or \textit{PL} are not on the body-fixed $z$-axis.

\subsection{Modeling} \label{sec:4_modeling}
\begin{figure}
    \centering
    \includegraphics[width=0.55\linewidth]{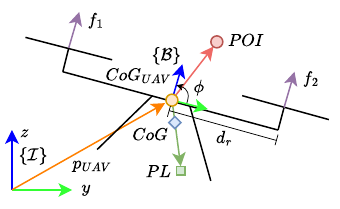}
    \vspace{-10pt}
    \caption{Coordinates and relevant positions for the planar UAV model in the $(y,z)$-plane}
    \label{fig:planarUAV}
    \vspace{-10pt}
\end{figure}

We first derive the dynamic model of the planar representation of the multirotor UAV with a payload in the $(y,z)$-plane with rotation $\varphi$ about the $x$-axis.
Let us therefore define the rotation matrix that maps points from the body-fixed frame $\mathcal{B}$ to the inertial frame $\mathcal{I}$ (as in Fig. \ref{fig:planarUAV}), the skew-symmetric matrix function $\bar{S}(a)$, and unit vectors $\bar{\bm{e}}_1$ and $\bar{\bm{e}}_2$ as
\begin{gather*}
    \bar{R} = \begin{bmatrix}
        \cos(\varphi) & -\sin(\varphi) \\
        \sin(\varphi) & \cos(\varphi)
    \end{bmatrix}, \quad \bar{S}(a) = \begin{bmatrix}
        0 & -a \\ a & 0
    \end{bmatrix}, \\ \quad \bar{\bm{e}}_1 = \begin{bmatrix}
        1 \\ 0
    \end{bmatrix}, \quad \bar{\bm{e}}_2 = \begin{bmatrix}
        0 \\ 1
    \end{bmatrix}, \quad \bar{J}= \begin{bmatrix}
        0 & -1 \\ 1 & 0
    \end{bmatrix}.
\end{gather*}
For the planar representation of the multirotor UAV system with a heavy payload, we again consider the four relevant geometric points \textit{CoG}$_{UAV}$, \textit{PL}, \textit{CoG}, and \textit{POI} as before, shown in Fig. \ref{fig:planarUAV}.
The sum of the inertia of the drone and the inertia of the payload about $x$ is $H_{\text{tot},xx}$.
We define $\bar{\bm{r}}_{PL}$ and $\bar{\bm{r}}_{POI}$ as the position of the payload and the point of interest, respectively, given by 
\begin{gather*}
    \bar{\bm{r}}_{PL} = \begin{bmatrix}
        y_{PL} \\ z_{PL}
    \end{bmatrix}, \quad
    \bar{\bm{r}}_{POI} = \begin{bmatrix}
        y_{POI} \\ z_{POI}
    \end{bmatrix}.
\end{gather*}
The positions of the payload $\bar{\bm{p}}_{PL}$ and point of interest $\bar{\bm{p}}_{POI}$ in the inertial frame $\mathcal{I}$ are defined as $\bar{\bm{p}}_{PL} = \bar{\bm{p}}_{UAV} + \bar{R} \bar{\bm{r}}_{PL}$, $\bar{\bm{p}}_{POI} = \bar{\bm{p}}_{UAV} + \bar{R} \bar{\bm{r}}_{POI}$.
Using the parallel axis theorem \cite{goldstein_classical_2007}, the total inertia about the $x$-axis is given by
\begin{gather*}
    \bar{H}_{CoG} = H_{tot,xx} + \frac{m_{UAV} m_{PL}}{m_{tot}}(y_{PL}^2+z_{PL}^2).
\end{gather*}
Finally, the positions of the total center of gravity of the UAV with payload are given by $\bar{\bm{p}}_{CoG} = \bar{\bm{p}}_{POI} - \bar{R} \hat{\bar{\bm{r}}}$, where $ \hat{\bar{\bm{r}}} = \bar{\bm{r}}_{POI} - \frac{m_{PL}}{m_{tot}} \bar{\bm{r}}_{PL}$.
For deriving the dynamic model of the UAV with payload using the Euler-Lagrange approach, the kinetic and potential energy functions are given by
\begin{gather*}
\begin{aligned}
    \mathcal{K} =& \frac{1}{2} m_{\text{tot}} \dot{\bar{\bm{p}}}_{CoG}^\top\dot{\bar{\bm{p}}}_{CoG} + \frac{1}{2} \dot{\varphi}^\top \bar{H}_{CoG}(\bar{\bm{r}}_{PL}) \dot{\varphi}, \\
    \mathcal{U} =& m_{\text{tot}} g \bar{\bm{e}}_2^\top \bar{\bm{p}}_{CoG}.
\end{aligned}    
\end{gather*}
Considering the Euler-Lagrange equation for each generalized coordinates in $\bm{\xi}=\begin{bmatrix}
    \bar{\bm{p}}_{POI} & \varphi
\end{bmatrix}^\top$:
\begin{gather*}
    \frac{d}{dt}\frac{\partial\mathcal{L}}{\partial \dot{\xi}_i} - \frac{\partial \mathcal{L}}{\partial \xi_i} = u_i,
\end{gather*}
yielding the dynamic equation for the planar system
\begin{gather}
\label{eq:dynamic_equations}
    \bar{B}(\varphi) \ddot{\bm{\xi}} + \bar{C}(\varphi, \dot{\varphi}) \dot{\bm{\xi}} + \bar{G}(\varphi) = \bar{M}(\varphi) \begin{bmatrix}
        T \\ \tau_{\varphi}
    \end{bmatrix}, \\
    \bar{B}(\varphi) = \begin{bmatrix}
        m_{tot} I_2 & -m_{tot}\bar{R}\bar{J}\hat{\bar{\bm{r}}} \\
        -m_{tot}\bar{R}\bar{J}\hat{\bar{\bm{r}}} & \bar{H}_{CoG}(\bar{\bm{r}}_{PL}) + m_{tot}\hat{\bar{\bm{r}}}^\top\hat{\bar{\bm{r}}}
    \end{bmatrix}, \\
    \bar{C}(\varphi, \dot{\varphi}) = \begin{bmatrix}
        0 & \dot{\varphi} R \hat{\bar{\bm{r}}} \\ 0 & 0
    \end{bmatrix}, \quad
    \bar{G}(\varphi) = \begin{bmatrix}
        m_{tot} g \bar{\bm{e}}_2 \\ -m_{tot}g\bar{\bm{e}}_2^\top R J \hat{\bar{\bm{r}}}
    \end{bmatrix}, \\
    \bar{M}(\varphi) = \begin{bmatrix}
        R \bar{\bm{e}}_2 & 0 \\ 0 & 1
    \end{bmatrix}.
\end{gather}

\subsection{Input Limits} \label{sec:4_thrustlimits}
Most UAVs do not allow negative thrusts and have a maximum rotor speed, resulting in a maximum individual thrust force.
To take these thrust limits into account, let us first define the relation between the individual motor thrust forces $f_1$ and $f_2$ and the total thrust $T$ and torque $\tau_{\varphi}$:
\begin{equation*}
    \begin{bmatrix}
        T \\
        \tau_{\varphi}
    \end{bmatrix} = \begin{bmatrix}
        f_1 + f_2 \\
        (d_r - y_{POI}) f_2 - (d_r + y_{POI}) f_1
    \end{bmatrix},
\end{equation*}
where $d_r$ is the distance along the body-fixed $y$-axis between the \textit{CoG}$_{UAV}$ and the thrust force $f_1$ and $f_2$, as seen in \mbox{Fig. \ref{fig:planarUAV}}.
From \eqref{eq:dynamic_equations}, we can derive that the equilibrium point $(\varphi, \dot{\varphi})=(0, 0)$ has an equilibrium thrust and torque given by $T_{\text{eq}} = m_{\text{tot}}g$ and $\tau_{\varphi, \text{eq}} = -m_{\text{tot}}g \bar{\bm{e}}_1^\top \hat{\bar{\bm{r}}}$.
The individual thrust forces are then given by $f_{1,\text{eq}} = \frac{g}{2 d_r} \left( m_{\text{tot}} d_r - m_{PL} y_{PL} \right)$ and $f_{2,\text{eq}} = \frac{g}{2 d_r} \left( m_{\text{tot}} d_r + m_{PL} y_{PL} \right)$.
From this, it is clear that the horizontal distance of the payload is limited, resulting in a maximum horizontal payload distance $|d_{{PL}, \text{max}}|$:
\begin{equation*}
    d_{{PL}, \text{max}} = \min\left(\frac{d_r}{m_{PL}} m_{\text{tot}}, \frac{d_r}{m_{PL}} \left| 2 \frac{f_{\text{max}}}{g} - m_{\text{tot}} \right| \right).
\end{equation*}
Here, $f_{\text{max}}$ is the maximum individual thrust force.
Therefore, $y_{PL}$ should satisfy
\begin{equation}
\label{eq:PLlim}
    |y_{PL}| < d_{{PL}, \text{max}}.
\end{equation}
A logical consequence is that the position of the \textit{CoG} cannot take values beyond the horizontal rotor position.

\subsection{Stability Analysis of the Planar System} \label{sec:4_stability}
Since the objective is to control the position of the \textit{POI} on the UAV, we perform input-output linearization to derive the corresponding zero dynamics.
To this end, we rewrite \eqref{eq:dynamic_equations} as
\begin{gather*}
\begin{aligned}
    \begin{bmatrix}
        \ddot{\bar{\bm{p}}}_{POI} \\ \ddot{\varphi}
    \end{bmatrix} =& \bar{B}(\varphi)^{-1} \bar{M}(\varphi) \begin{bmatrix}
        T \\ \tau_{\varphi}
    \end{bmatrix} - \bar{B}(\varphi)^{-1} \bar{M}_{CG}, \\
    \ddot{\bar{\bm{p}}}_{POI} =& \bar{U}(\varphi) \bar{M}(\varphi) \begin{bmatrix}
        T \\ \tau_{\varphi}
    \end{bmatrix} -\bar{U}(\varphi) \bar{M}_{CG},
\end{aligned}
\end{gather*}
with
\begin{gather*}
\begin{aligned}
    &\bar{M}_{CG} = \bar{C}(\varphi, \dot{\varphi}) \begin{bmatrix}
        \dot{\bar{\bm{p}}}_{POI} \\ \dot{\varphi}
    \end{bmatrix} + \bar{G}(\varphi), \\
    &\bar{U}(\varphi) = \begin{bmatrix}
        1 & 0 & 0 \\ 0 & 1 & 0
    \end{bmatrix} \bar{B}(\varphi)^{-1}.
\end{aligned}
\end{gather*}
The zero dynamics inputs when controlling $\ddot{\bar{\bm{p}}}_{POI}$ to zero are defined as
\begin{gather*}
\begin{aligned}
    &\begin{bmatrix}
        T_{zd} \\ {\tau_{\varphi}}_{zd}
    \end{bmatrix} =
    \left( \bar{U}(\varphi) \bar{M}(\varphi) \right)^{-1} \bar{U}(\varphi) \bar{M}_{CG}. 
    %
\end{aligned}
\end{gather*}
%
Substituting the zero-dynamics input yields \begin{equation} \label{eq:zerodyn2D} \ddot{\varphi} = \frac{ -\dot{\varphi}^{2}\bar{\bm e}_{1}^{\top}\hat{\bar{\bm r}} - m_{\mathrm{tot}}g\sin\varphi }{ m_{\mathrm{tot}}\bar{\bm e}_{2}^{\top}\hat{\bar{\bm r}} }. \end{equation}
If we recall the definition of $\alpha = \bar{\bm{e}}_2^\top\hat{\bar{\bm{r}}} = z_{POI} - \frac{m_{PL}}{m_{\text{tot}}} z_{PL}$, we can then generalize the zero dynamics in \eqref{eq:zerodyn2D} to
\begin{equation}
\label{eq:xysystem}
    \begin{cases}
    \dot{x} = y,\\[4pt]
    \dot{y} = -a y^2 - b \sin x,
    \end{cases}
\end{equation}
where $a = \frac{\bar{\bm{e}}_1^\top \hat{\bar{\bm{r}}}}{m_{\text{tot}}\alpha}$, $b = \frac{g}{\alpha}$, and where $x$ represents $\varphi$ and $y$ represents $\dot{\varphi}$.
We now proceed to prove that the system above is Lyapunov stable for $\alpha>0$, i.e., the \textit{POI} should be above the \textit{CoG}.
For this, we first state Theorem \ref{the:xy}, which analyses the stability of \eqref{eq:xysystem} for general $a$ and $b$.
\begin{theorem}
\label{the:xy}
    The origin of the planar system given by \eqref{eq:xysystem}, for general $a, b\in\mathbb{R}$, is locally Lyapunov stable if $b>0$.
\end{theorem}
\begin{proof}
A Lyapunov function candidate of the system in \eqref{eq:xysystem} is given by
\[
H(x,y)
= \frac{1}{2} e^{2ax} y^{2}
\;+\;
\frac{b}{4a^{2}+1}\Big( e^{2ax}(2a\sin x - \cos x) + 1 \Big),
\]
which is positive in $[-\pi, \pi]$. Its time derivative is given by
\begin{gather*}
\begin{aligned}
    \dot{H}(x,y) &= e^{2ax} (a \dot{x} y^2 + y \dot{y}) + e^{2ax} b \dot{x} \sin x = 0.
\end{aligned}
\end{gather*}
Thus, $H(x,y)$ is indeed constant along trajectories.
The invariant level sets of $H(x,y)$ are illustrated in Fig. \ref{fig:phase}, for given values of $a$ and $b$.
\begin{figure}
    \centering
    \includegraphics[width=0.85\linewidth]{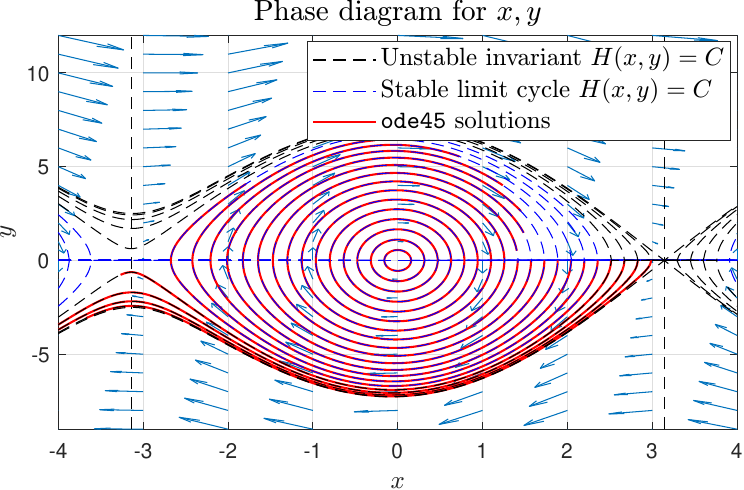}
    \vspace{-5pt}
    \caption{Phase diagram of the zero dynamics for $x,\,y$ for $a=0.0179$ and $b=12.4855$, where the analytical level sets correspond with the numerical results.
    Unstable level sets are present for $H(x_0, y_0) < \frac{b}{4a^2 + 1} (1 + e^{-2 |a| \pi})$.}
    \label{fig:phase}
    \vspace{-10pt}
\end{figure}
The gradient of $H(x,y)$ evaluated at the origin is zero and the Hessian of $H(x,y)$ is given by
\[
\nabla^2 H(x,y)
=
e^{2ax}
\begin{bmatrix}
2a^2 y^2 + 2ab\sin x + b\cos x & 2ay\\
2ay & 1
\end{bmatrix}.
\]
which, evaluated at the origin, is a positive definite function when $b>0$.
Thus, we can pick a local domain as a sufficiently small neighborhood of the origin in which $H(x,y)$ is zero at the origin, positive definite, and the derivative along trajectories is zero.
Applying~\cite[Th. 4.1]{khalil_nonlinear_2002}, we can then conclude Lyapunov stability of the origin for $b>0$.
\end{proof}

\begin{theorem}
\label{the:2limit}
    The zero dynamics of the planar UAV with payload system given in \eqref{eq:zerodyn2D} is Lyapunov stable if $\alpha > 0$ and if $y_{PL}$ satisfies \eqref{eq:PLlim}.
\end{theorem}

\begin{proof}
    We can rewrite the system in \eqref{eq:zerodyn2D} to the general form given in \eqref{eq:xysystem}, for $x = \varphi$, $y = \dot{\varphi}$ $a = \frac{\bar{\bm{e}}_1^\top \hat{\bar{\bm{r}}}}{m_{\text{tot}}\bar{\bm{e}}_2^\top \hat{\bar{\bm{r}}}}$, and $b = \frac{g}{\alpha}$.
    Given Theorem \ref{the:xy}, the equilibrium of the system ($\varphi = \dot{\varphi} = 0$) is then Lyapunov stable for $b > 0$, which is true for $\alpha > 0$.
    Furthermore, if $y_{PL}$ satisfies \eqref{eq:PLlim}, we know that the corresponding zero dynamic inputs $T_{zd}$ and ${\tau_{\varphi}}_{zd}$ at the equilibrium point result in individual thrust forces $f_1$ and $f_2$ such that $0 < f_i < f_\text{max}$ $\forall i \in \{1, 2\}$.
\end{proof}

\begin{remark}
    Although the horizontal distance of the payload is limited due to motor thrust force limits, the horizontal position of the payload and/or \textit{POI} does not influence the stability of the system given in \eqref{eq:zerodyn2D}.
\end{remark}

%
\section{ILLUSTRATIVE EXAMPLES} \label{sec:simulations}
In this section, we numerically verify the results from Section \ref{sec:3D_proof} in Section \ref{sec:sim_3D_onz}.
We then continue to give insight into the stability of the full system with simulations for when the payload is not on the body-fixed $z$-axis in Section \ref{sec:sim_3D_not_onz}.

\par Simulations have been performed in MATLAB using the \texttt{ode45} solver, where the control law in \eqref{eq:zd_control_law} has been applied to keep the position of the \textit{POI} fixed at the origin.
A relative tolerance of $10^{-5}$, an absolute tolerance of $10^{-6}$, and a maximum step size of $0.001$ [$s$] have been used.
The used system masses and inertias are $m_{UAV} = 22.0 [kg]$, $m_{PL} = 6.0 [kg]$, $H_{\text{tot},xx} = H_{\text{tot},yy} = 0.25\, [kg\cdot m^2]$ and $H_{\text{tot},zz} = 0.14\, [kg\cdot m^2]$.
Furthermore, for all simulations, $\bm{\eta}_0 = \begin{bmatrix}
    0.8 & 0.3 & 0
\end{bmatrix}^\top$ and $\dot{\bm{\eta}}_0 = \begin{bmatrix}
    2.0 & -1.0 & 0
\end{bmatrix}^\top$.

\subsection{3D simulations with payload on the body-fixed \texorpdfstring{$z$}{z}-axis} \label{sec:sim_3D_onz}
\begin{figure}
    \centering
    \includegraphics[width=0.9\linewidth]{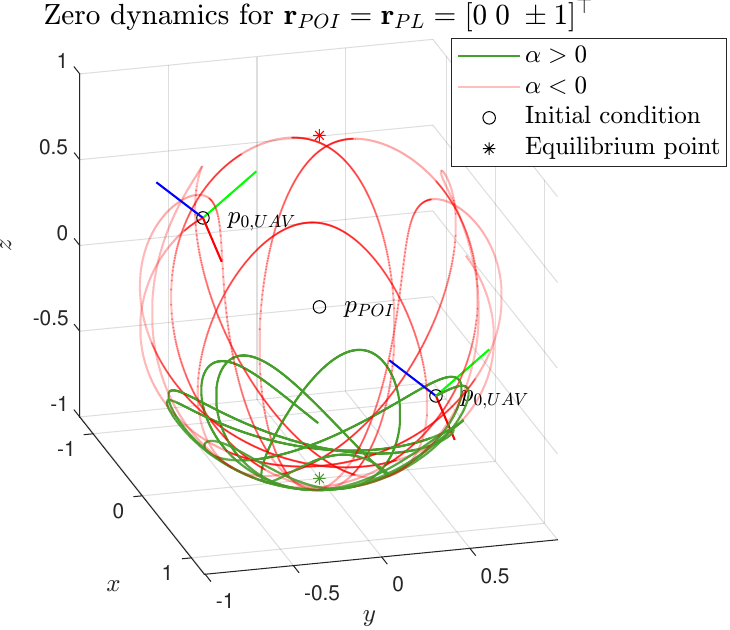}
    \vspace{-5pt}
    \caption{Evolution of the zero dynamics trajectories of $\bm{p}_{UAV}$.
    The trajectory for $\alpha>0$ orbits the equilibrium point; the trajectory for $\alpha<0$ does not.}
    \label{fig:zerodyn3d_onz}
    \vspace{-5pt}
\end{figure}
\begin{figure}
    \centering
    \includegraphics[width=0.95\linewidth]{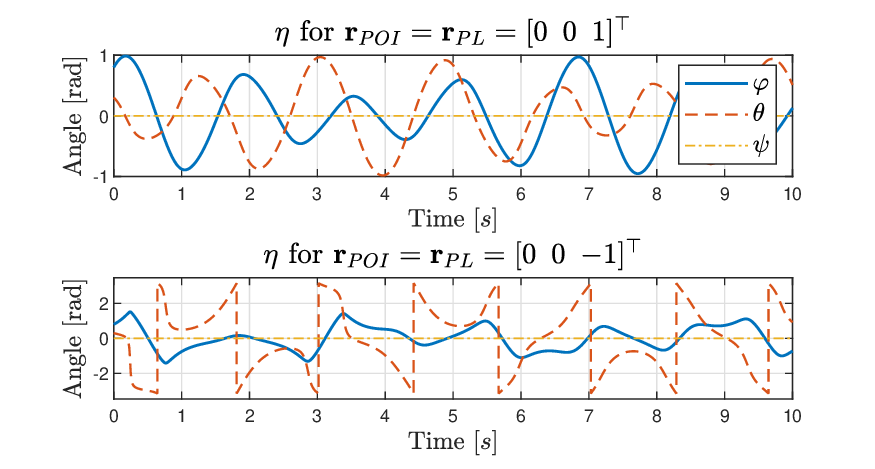}
    \vspace{-5pt}
    \caption{Evolution of the Euler angles $\varphi$, $\theta$, and $\psi$ over time, wrapped between $-\pi$ and $\pi$.}
    \label{fig:angles_over_time_onz}
    \vspace{-10pt}
\end{figure}
\begin{figure}
    \centering
    \includegraphics[width=0.95\linewidth]{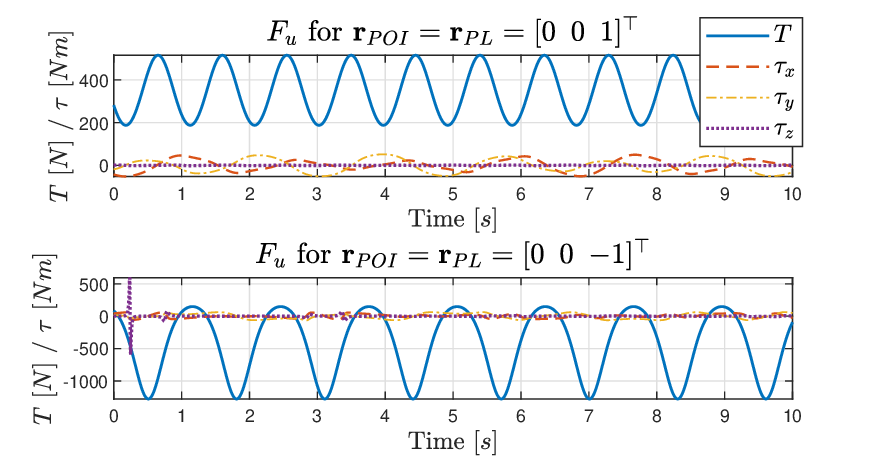}
    \vspace{-5pt}
    \caption{Control inputs $F_u$ over time}
    \label{fig:inputs_onz}
\end{figure}
In Section \ref{sec:3D_proof}, we proved that the multirotor UAV with payload is only stable if the \textit{POI} is above the total \textit{CoG}.
This stability proof holds when the \textit{PL} is on the body-fixed $z$-axis, and this stability can also be seen in Fig. \ref{fig:zerodyn3d_onz}, where stable oscillations are observed around the equilibrium point only when $\alpha>0$.
Note that in this figure, the $POI=PL$ is fixed at the origin, and the trajectory of the center position of the UAV $\bm{p}_{UAV}$ is visualized.
The evolution of the Euler angles is visualized in Fig. \ref{fig:angles_over_time_onz} and the corresponding zero dynamics inputs are shown in Fig. \ref{fig:inputs_onz}.
Here, we also see that the thrust input is negative when $\alpha<0$, which is infeasible in practice.


\subsection{3D simulations with payload not on the body-fixed \texorpdfstring{$z$}{z}-axis} \label{sec:sim_3D_not_onz}
\begin{figure}
    \centering
    \includegraphics[width=0.9\linewidth]{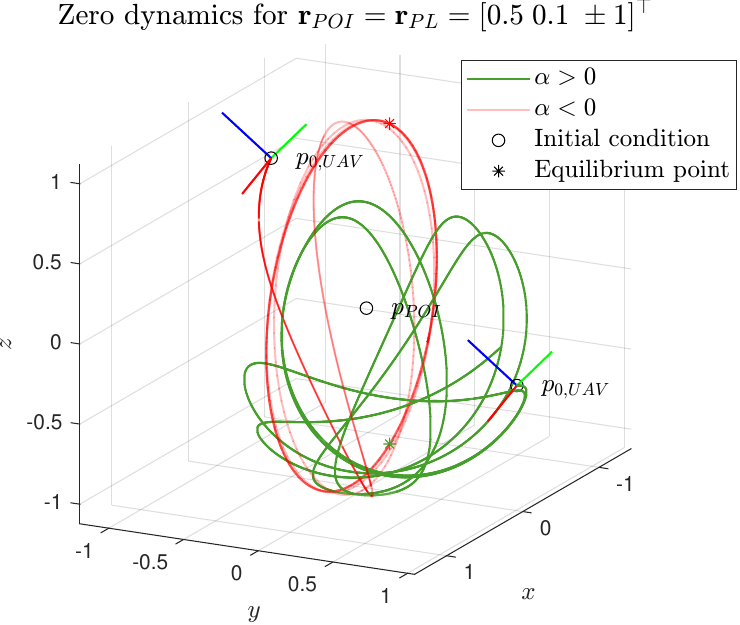}
    \vspace{-5pt}
    \caption{Evolution of the zero dynamics trajectories of $\bm{p}_{UAV}$.
    The trajectory for $\alpha>0$ orbits the equilibrium point; the trajectory for $\alpha<0$ blows up.}
    \label{fig:zerodyn3d}
    \vspace{-10pt}
\end{figure}
\begin{figure}
    \centering
    \includegraphics[width=0.95\linewidth]{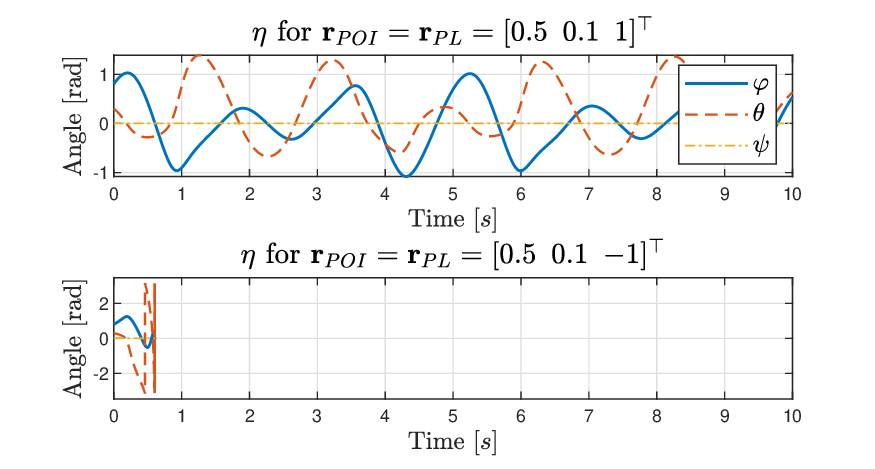}
    \vspace{-5pt}
    \caption{Evolution of the Euler angles $\varphi$, $\theta$, and $\psi$ over time, wrapped between $-\pi$ and $\pi$.}
    \label{fig:angles_over_time}
\end{figure}
Since it is intractable to investigate the stability of a three-dimensional UAV system with a heavy payload, simulations are performed to get insight into how the horizontal position of the payload influences the stability of the system.
Fig. \ref{fig:angles_over_time} shows typical system behavior, where the system is only Lyapunov stable if $\alpha>0$, showing oscillating behavior.
A 3D visualization of the evolution of the UAV rotation can be seen in Fig. \ref{fig:zerodyn3d}.
The system was also simulated for a grid of values of $\varphi_0$ and $\theta_0$ as the only nonzero initial conditions, ranging between $-\frac{\pi}{6}$ and $\frac{\pi}{6}$.
Here, stable oscillations were shown as long as $\alpha>0$, emphasising the hypothesis that the system is also Lyapunov stable when the \textit{POI} and \textit{PL} are not on the body-fixed $z$-axis.

\section{CONCLUSION} \label{sec:conclusion}
This paper shows that, when controlling an arbitrary point of interest on a multirotor UAV with a rigidly attached heavy payload, it is better to place the point of interest above the UAV's overall center of gravity.
Therefore, if the payload has to be stabilized, it is better to place it above the UAV, rather than below it.
These insights can be used as a design guide when attaching (multiple) payloads and sensors to a multirotor UAV.
Future work includes analyzing whether the conclusions still hold under disturbances and performing experiments with a multirotor UAV system with a heavy payload to investigate whether the analytical results transfer to a real-world application.

\addtolength{\textheight}{-12cm}   




\bibliographystyle{IEEEtran}
\bibliography{IEEEabrv,references_no_url}

\end{document}